\documentclass[sigconf,nonacm]{acmart}
\copyrightyear{2024}
\acmYear{2024}
\setcopyright{rightsretained}
\acmConference[DAC '24]{61st ACM/IEEE Design Automation Conference}{June 23--27, 2024}{San Francisco, CA, USA}
\acmBooktitle{61st ACM/IEEE Design Automation Conference (DAC '24), June 23--27, 2024, San Francisco, CA, USA}
\acmDOI{10.1145/3649329.3657377}
\acmISBN{979-8-4007-0601-1/24/06}
\acmConference[DAC '24]{Design Automation Conference 2024}{June 2024}{San Francisco, CA, USA}
\acmPrice{15.00}
\acmISBN{978-x-xxxx-xxxx-x/xx/xx}
\acmDOI{10.1145/nnnnnnn.nnnnnnn}
\usepackage{amsmath,amsfonts,amsthm}

\usepackage[ruled]{algorithm}
\usepackage[noend]{algpseudocode}
\usepackage[normalem]{ulem}
\usepackage{graphicx}
\usepackage{textcomp}
\usepackage{xcolor}
\usepackage{color}
\usepackage{etoolbox}
\usepackage{subcaption}
\usepackage{tabularx}
\usepackage{multirow}
\usepackage{makecell}
\usepackage{textcomp}
\usepackage{tabularx}
\usepackage{booktabs}
\usepackage{makecell}

\usepackage{amsmath,amsfonts,bm}

\def\eqref#1{equation~\ref{#1}}

\def\1{\bm{1}}

\usepackage{hyperref}
\usepackage{url}
\usepackage[english]{babel}

\usepackage{booktabs}
\AtBeginDocument{%
  \providecommand\BibTeX{{%
    \normalfont B\kern-0.5em{\scshape i\kern-0.25em b}\kern-0.8em\TeX}}}

\newcommand{\OURS}{\textsc{EPIM}}

\newcommand{\Tau}{\tau}

\begin{document}

\title{EPIM: Efficient Processing-In-Memory Accelerators based on Epitome}

\author{Chenyu~Wang$^{1*}$,Zhen~Dong$^{2*\dagger}$,Daquan~Zhou$^{3*\dagger}$,Zhenhua~Zhu$^{1}$,Yu~Wang$^{1\dagger}$,Jiashi~Feng$^{3}$,Kurt~Keutzer$^{2}$}

\affiliation{
\vspace{0em}
\institution{$^{1}$Tsinghua University \hspace{0.3em}$^{2}$UC Berkeley \hspace{0.3em}$^{3}$ByteDance}
\country{}
}

\affiliation{
\small
\institution{$^*$ Equal contribution \hspace{0.3em}$^\dagger$ Corresponding to: zhendong@berkeley.edu, zhoudaquan21@gmail.com, yu-wang@mail.tsinghua.edu.cn }
\country{}
\vspace{10mm}
}

\begin{abstract}
 %
The utilization of large-scale neural networks on Processing-In-Memory (PIM) accelerators encounters challenges due to constrained on-chip memory capacity.
To tackle this issue, current works explore model compression algorithms to reduce the size of Convolutional Neural Networks (CNNs). Most of these algorithms either aim to represent neural operators with reduced-size parameters (e.g., quantization) or search for the best combinations of neural operators (e.g., neural architecture search). Designing neural operators to align with PIM accelerators' specifications is an area that warrants further study.
In this paper, we introduce the \underline{E}pitome, a lightweight neural operator offering convolution-like functionality, to craft memory-efficient CNN operators for \underline{PIM} accelerators (\OURS).
On the software side, we evaluate epitomes' latency and energy on PIM accelerators and introduce a PIM-aware layer-wise design method to enhance their hardware efficiency. We apply epitome-aware quantization to further reduce the size
of epitomes.
On the hardware side, we modify the datapath of current PIM accelerators to accommodate epitomes and implement a feature map reuse technique to reduce computation cost.
Experimental results reveal that our 3-bit quantized \OURS-ResNet50 attains 71.59\% top-1 accuracy on ImageNet, reducing crossbar areas by 30.65$\times$. EPIM surpasses the state-of-the-art pruning methods on PIM. 

\end{abstract}

\maketitle

\pagestyle{plain}

\section{Introduction}
\label{sec:intro}

The emergence of Processing-In-Memory (PIM) accelerators offers a promising approach to boost the computational efficiency of neural network operations~\cite{chi2016prime}. Leveraging the capability to conduct computations directly within memory and eliminating the need for frequent data transfers between memory and computing units, PIM accelerators offer a substantial improvement in energy efficiency. Remarkably, these accelerators surpass the performance of GPU and CMOS ASIC solutions by more than 100-fold~\cite{shafiee2016isaac,chi2016prime}. 

Despite the advances in current PIM accelerators, they still face certain challenges. One challenge arises from the higher cost of writing to emerging memory compared to reading from it. Due to this limitation, PIM accelerators typically require loading all neural network weights onto memristor crossbars prior to conducting computations. However, modern large-scale neural network models have seen a significant increase in parameter sizes. Even for the existing PIM accelerators with maximum memristors,  there still exists a significant challenge in deploying a standard backbone model like ResNet-50\cite{wan2022compute}. As such, the deployment of advanced Convolutional Neural Networks (CNN) models on PIM is hindered. 
 
To address this challenge, researchers have proposed methods that fall into two main categories. Various methods, including pruning and quantization, aim to represent neural operator weights using a reduced number of parameters\cite{sun2020energy,chu2020pim}.
Conversely, approaches such as neural architecture search concentrates on identifying optimal combinations of neural operators\cite{jiang2020device,sun2022gibbon}. 
Despite the advancements, the majority of these studies have focused on adapting existing convolution operators. The crucial endeavor of developing compact and high-performance operators specifically tailored for PIM accelerators has largely been overlooked.   \par

Designing an operator for PIM requires meeting essential standards. For an effective PIM-specific neural operator, it must address memristor limitations by using fewer memristors than current convolution operators without sacrificing accuracy. Additionally, the operator should require minimal adjustments to existing software and PIM accelerators, ensuring an efficient and seamless transition to the new design.\par

To craft such neural operator, we began by reviewing existing literature. Prior research highlights the potential of leveraging overlaps in convolution kernels, introducing a compact neural operator termed \textquotedblleft epitome\textquotedblright \cite{zhou2019neural}. This operator reconstructs convolution kernels from a smaller epitome tensor. A more compact neural network can be crafted by replacing the its convolutions with epitomes. We identified that, with software and hardware adjustments, epitome could be adapted to PIM accelerators. Specifically, from the hardware side, epitome only introduces minor modifications to the data path. Hence, we adopted epitome as our foundation. By integrating epitome with our proposed methods, we achieved notable compression rates while maintaining a high accuracy.

Our contributions can be summarized as follows:
\begin{itemize}
    \item We pioneer the utilization of epitome as an efficient neural operator for PIM accelerators, which is a distinct perspective compared to existing model compression approaches. 
    \item We recognize the correlation between epitome shape and both latency and energy consumption. Furthermore, we implement a PIM-aware layer-wise design method to optimize the epitome shape.
    \item We design three modules in the data path—IFRT, IFAT, and OFAT—to facilitate the deployment of epitome operators. We design a feature map reuse technique termed as \textquotedblleft Channel Wrapping\textquotedblright\ to further reduce computation cost.
    \item We propose an epitome-aware quantization method. This approach improves accuracy for epitomes under ultra-low bit quantization.
\end{itemize}

Our proposed methods yield outstanding results. Our 3-bit quantized \OURS-ResNet50 model demonstrates remarkable performance by achieving an accuracy of 71.59\% on the challenging ImageNet dataset with a crossbar compression rate of 30.65$\times$.
\section{Related work}
\label{sec:related}

\subsection{Process-In-Memory Architecture for CNN Acceleration}

Commonly, PIM architecture consists of multiple memristor crossbars. The weight matrices of CNN are mapped as the conductance of the memristor crossbar cells. However, the writing latency of the memristor crossbar cell is multiple times larger than the reading latency\cite{chi2016prime}. Before inference, the weight matrices of all CNN layers need to be mapped onto memristor crossbars. Such restrictions stimulate the need to support model compression algorithms on PIM architecture. 


\begin{figure}[h]
\centering
\scalebox{1}{
\includegraphics[trim=60 30 50 30, clip, width=0.4\textwidth]{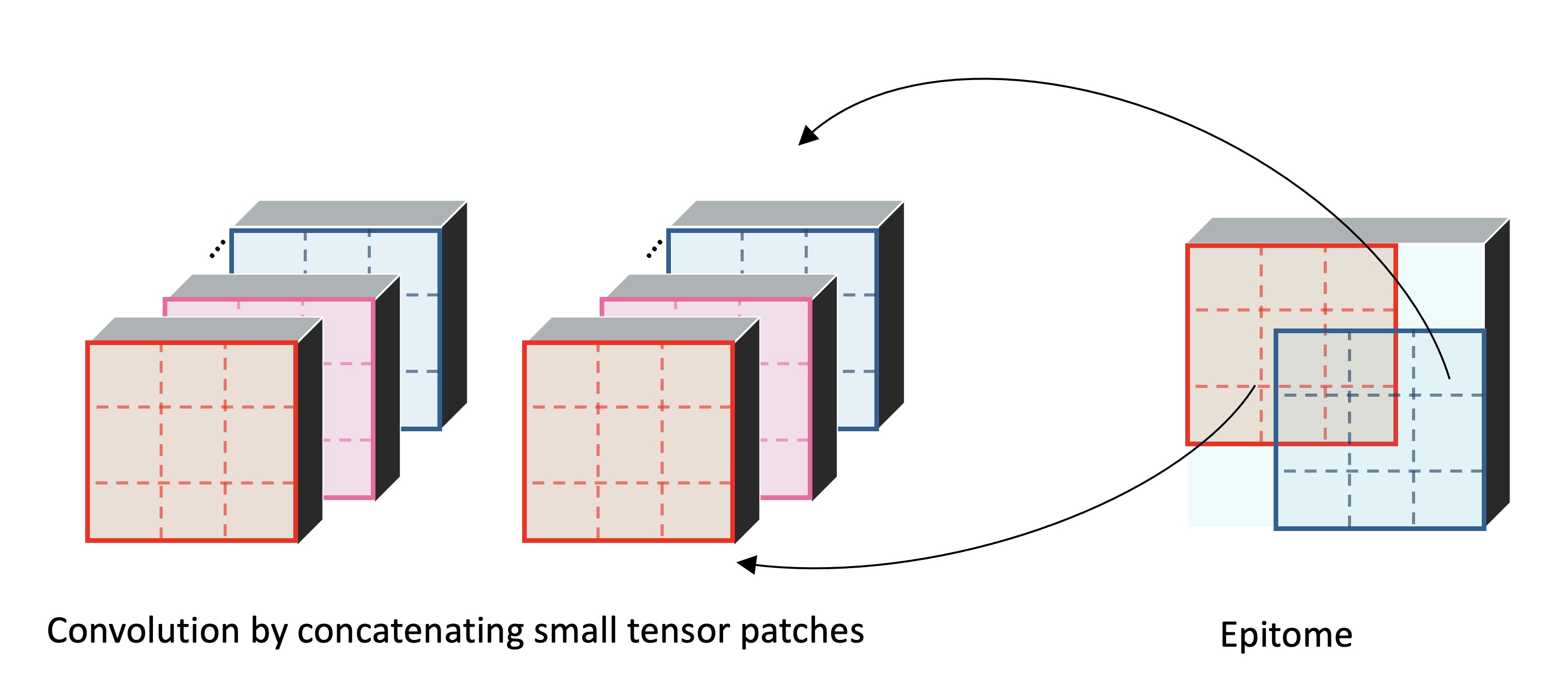}}
\caption{Epitome's Reconstruction into Convolution}

\label{fig:epim_illu}
\end{figure}

\subsection{Epitome}

%
The compact neural operator, epitome, represents the neural network with a compact set of parameters $E=\{E_i\}$, and a  sampler $\Tau$.
$\{E_i\}$ (epitomes) are a set of learnable parameters whose size are significantly smaller than the convolutions', and the sampler $\Tau$ repeatedly samples small patches from the epitomes. Each time of sampling, a sampler returns a small patch of epitome: 

\begin{equation}
\small
\label{eqn:sub_tensor_epitome}
    E_s = E_i[p:p+w, q:q+h, c_{in}:c_{in}+\beta_1, c_{out}:c_{out}+\beta_2],
\end{equation}
where $(p,q,c_{in},c_{out})$ denote the starting index of the sub-tensor in the epitome and $(w,h,\beta_1, \beta_2)$ denotes the length of the sub-tensor along each dimension. The sampling process is repeated until the fetched small pieces of weight tensors can be concatenated to match the dimensions of the convolution. Figure \ref{fig:epim_illu} is a visual illustration of such process. Unlike convolutions, whose shapes are determined by the input and output channels, epitomes have greater flexibility in shape. They can be of four-dimensional tensors with any size, as long as their sampled patches can reconstruct the convolutions\cite{zhou2019neural}.

\subsection{Quantization}

Quantization methods reduce model size by using low bitwidth for weights and activations in neural networks\cite{dong2019hawqv2}.
In quantization, the quantizer maps weights and activations into integers with a real-valued scaling factor $S$ and an integer zero point $Z$ as follows:
\begin{equation}
\label{eq:quantization_formula}
\small
Q(r) = \text{Int}\big({r}/{S}\big)-Z
\end{equation}
where $Q$ is the quantization operator, $r$ is a real-valued input, and $\text{Int()}$ represents a rounding operation. For a k-bit quantization, the scaling factor can be determined as follows:
\begin{equation}
\small
    S = \frac{\beta - \alpha}{2^{k} - 1}
\label{eq:quantization}
\end{equation}
$[\alpha, \beta]$ denotes the quantization range. A straightforward choice is to use the min/max of the signal for the clipping range, i.e., $\alpha=r_{min}$, and $\beta=r_{max}$.

\section{Framework Overview}
\label{sec:framework}


\begin{figure*}[ht]
\centering
\scalebox{1}{
\includegraphics[width=0.9\textwidth]{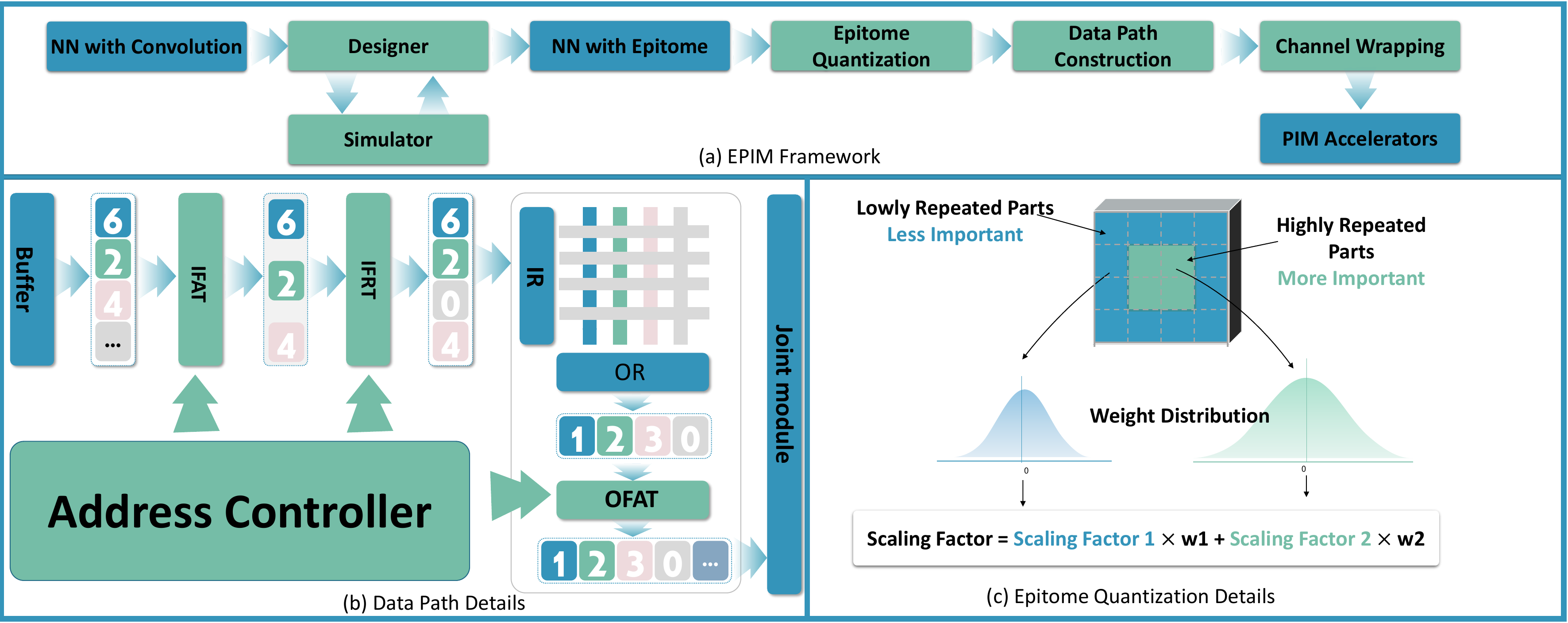}}
\caption{Illustration of EPIM Framework. }

\label{fig:main}
\end{figure*}

The general framework for designing and training neural networks with epitomes is presented in Figure \ref{fig:main}(a). EPIM begins with any convolution-based neural network. Subsequently, epitome designer is used to replace the convolutions by epitomes. This task can be performed manually or by using an evolutionary search to design a well-performing set of epitomes, informed by feedback from a simulator. Post-design, we attain a neural network where the operators are epitomes. After training, the epitome designer converts the floating point model to fixed-point. Then, we modify the data path and design the feature map reuse strategy, termed as \textquotedblleft Channel Wrapping\textquotedblright. After these steps, a well-crafted epitome based neural network can be deployed on PIM accelerators.

\section{EPIM Design Method}
\label{sec:method}

This section introduces our method \OURS. Section~\ref{subsec:epitome} shows the adjustments of epitome on PIM accelerators.
Section~\ref{subsec:quantization} presents details of our quantization method.
Section~\ref{subsec:pim} describes the specifications of our PIM accelerator.

\subsection{Mapping and Adjusting Epitomes for PIM Accelerators}
\label{subsec:epitome}


%
%
%
Just like convolutions, epitomes are four-dimensional tensors that can be mapped onto memristors in a similar manner. The distinction between epitomes and convolutions arises during inference: while all word lines and bit lines corresponding to a convolution's weight are activated all together, an epitome causes a memristor crossbar to be activated multiple times. Each activation only engages the word lines and bit lines associated with small patches sampled from the epitome. In our implementation, we use the same mapping strategy as \cite{zhu2020mnsim}.

Motivated by the size flexibility of the epitomes, we can adjust their shapes to better utilize memristors. Specifically, we aim for $c_{out}$ and $c_{in}\times p \times q$ to align as integral multiples of the crossbar size. Following the mapping approach from \cite{zhu2020mnsim}, where $c_{in}$, $p$ and $q$ dimensions correspond to word lines, while $c_{out}$ dimension corresponds to bit lines on memristor crossbars. By setting to these dimensions to desired values, we ensure epitomes make full use of memristor crossbars.

\subsection{Quantization Method for Epitome}
\label{subsec:quantization}

When directly quantizing the epitome with ultra-low precision, we observed a notable decrease in accuracy. Table \ref{tab:quantization_results} indicates that while the 5-bit quantized weights for ResNet-50's epitome achieve 73.59\% accuracy, this drops to 69.95\% for 3-bit weights. To enhance accuracy under such low-bit quantization, we suggest two adjustments:  Firstly, given the parallel computation between PIM accelerator crossbars, we allocate a scaling factor to each crossbar. Secondly, we observe that the epitome reconstructs convolution by sampling overlapping small patches, with certain patches from the epitome being repeated more often than others in the reconstructed convolution. When quantizing the epitome, it's logical to assign different importance levels to different parts based on their repetition frequency. As depicted in Figure \ref{fig:main}(c), we notice that the center parts (green) of epitomes are usually repeated more frequently than the other parts (blue). We calculate the weighted sum of the epitome's maximum and minimum values in the center parts and other parts as quantization range $\alpha$ and $\beta$, based on these repetitions. The detailed calculation process is as follows:

\begin{align}
\small
\alpha = w_{1} \times {(r_{min})}_{overlap} + w_{2} \times {(r_{min})}_{others} \\
\beta = w_{1} \times {(r_{max})}_{overlap} + w_{2} \times {(r_{max})}_{others} 
\end{align}

$w_{1}$ and $w_{2}$ are two hyperparameters denoting the significance of distinct sections of the epitome. We then can calculate scaling factor $S$ by Equation \ref{eq:quantization}.
A comprehensive analysis validating the proposed techniques can be found in ablation study.
\subsection{Architecture and Data Path of EPIM}

\label{subsec:pim}


The epitome method differs from traditional convolutions in that it breaks down the convolution into multiple smaller kernels. These kernels interact with specific sections of the input feature map to produce corresponding segments of the output feature map. To navigate this during EPIM computation, we need to pinpoint the relationships between the input data and the kernel storage addresses. To accommodate this, we introduce slight changes to the data-path of current PIM accelerators, integrating three index tables as visualized in Figure~\ref{fig:main}(b).

In practice, this means selecting the right inputs from the buffer and positioning them on the appropriate word lines of PIM crossbars. Also, voltages for those word lines connected to weights not part of this computation segment should be set to zero. Similarly, the outputs from the bit line must fit correctly within the output feature map. To efficiently manage this without adding to runtime, we've enhanced the base PIM accelerators with three index tables: the Input Feature Address Table (IFAT) and Input Feature Row Table (IFRT) for the input features, and the Output Feature Address Table (OFAT) for the outputs, as shown in Figure \ref{fig:main}(b). The remaining PIM accelerator components remain consistent with existing work \cite{zhu2019configurable}.

The IFAT comprises a sequence of start and stop index pairs. These indices pinpoint the position of the input feature that corresponds to the epitome kernel used in the current round, and the inputs stored between these indices are necessary for computation. The quantity of these index pairs equates to the number of times the memory device crossbars are activated. The IFRT contains multiple sequences used to identify the positions of word lines where the weights connected to this word line are used in the current round. The length of the sequence is equivalent to the number of PIM crossbar rows. The quantity of sequences should be equal to the number of small patches sampled from an epitome, given that we need to use it each time we load inputs. In the EPIM data-path, the process begins with the address controller generating a continuous address. After being processed by IFAT and IFRT, the input features are correctly transferred from the buffer into the input register. \par

Similar to the IFAT, the OFAT also consists of start and stop index pairs. These pairs indicate the position of the result within the entire output feature. The number of these index pairs corresponds to the number of small patches. After all small patches have been activated, and before the output features are forwarded to the next layer, the OFAT comes into play. It employs the joint module to reconstruct the output feature map. This is achieved by adding the output features with identical start and stop indices and concatenating those with sequential indices.

\vspace{-5mm}

\section{EPIM Design Optimization}
\label{sec:design_opt}

\begin{figure}[h]
\centering
\scalebox{1}{
\includegraphics[width=0.32\textwidth]{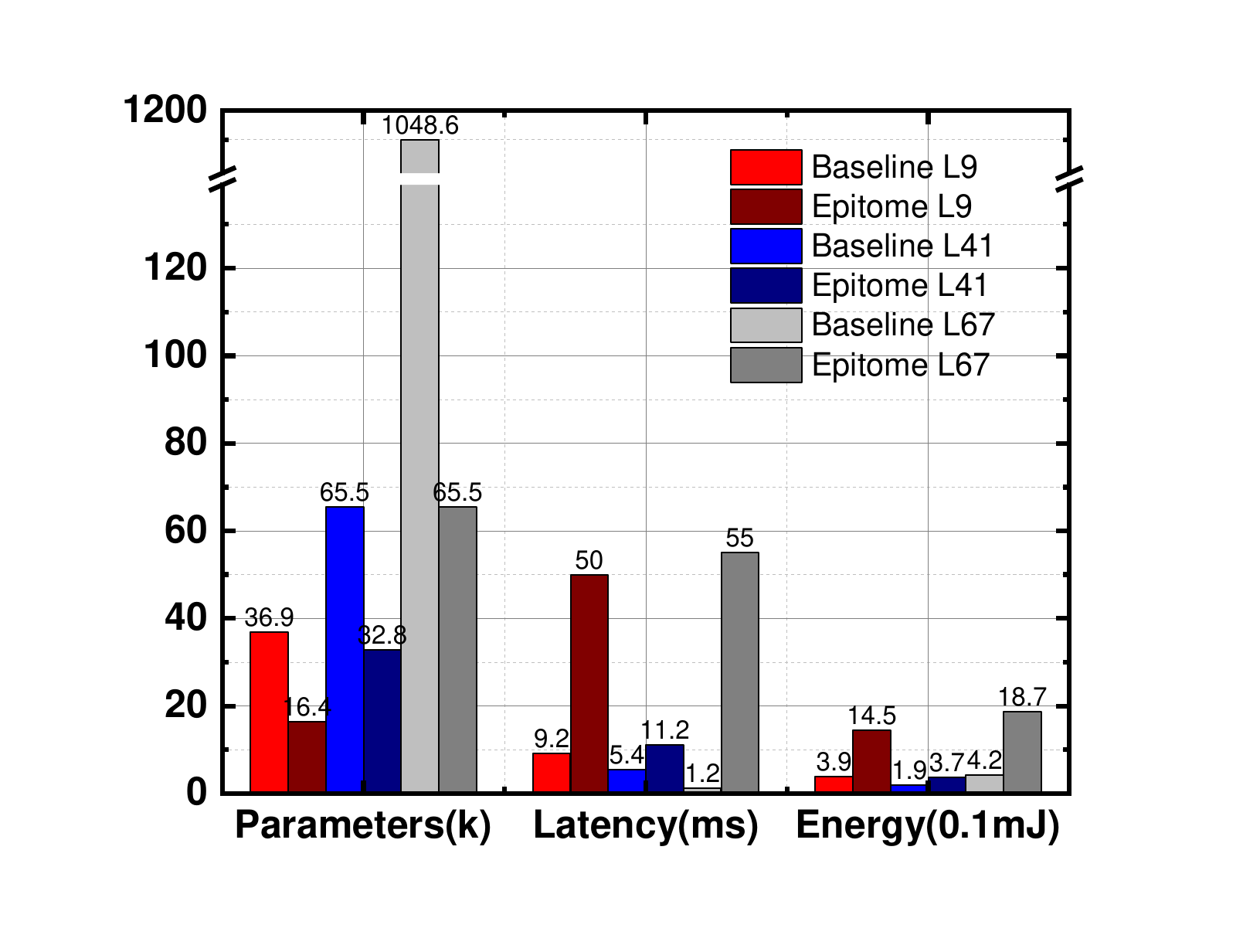}}
\caption{The parameter size, latency, and energy consumption correspond to different layers (Layer 9, 41, 67) in ResNet50, with or without epitome.}

\label{fig:opt_motivation}
\end{figure}

\subsection{Motivation}
\label{subsec:motivation}
\begin{figure*}[h]
\centering
\begin{subfigure}{0.32\textwidth}
    \centering
    \includegraphics[width=0.9\textwidth]{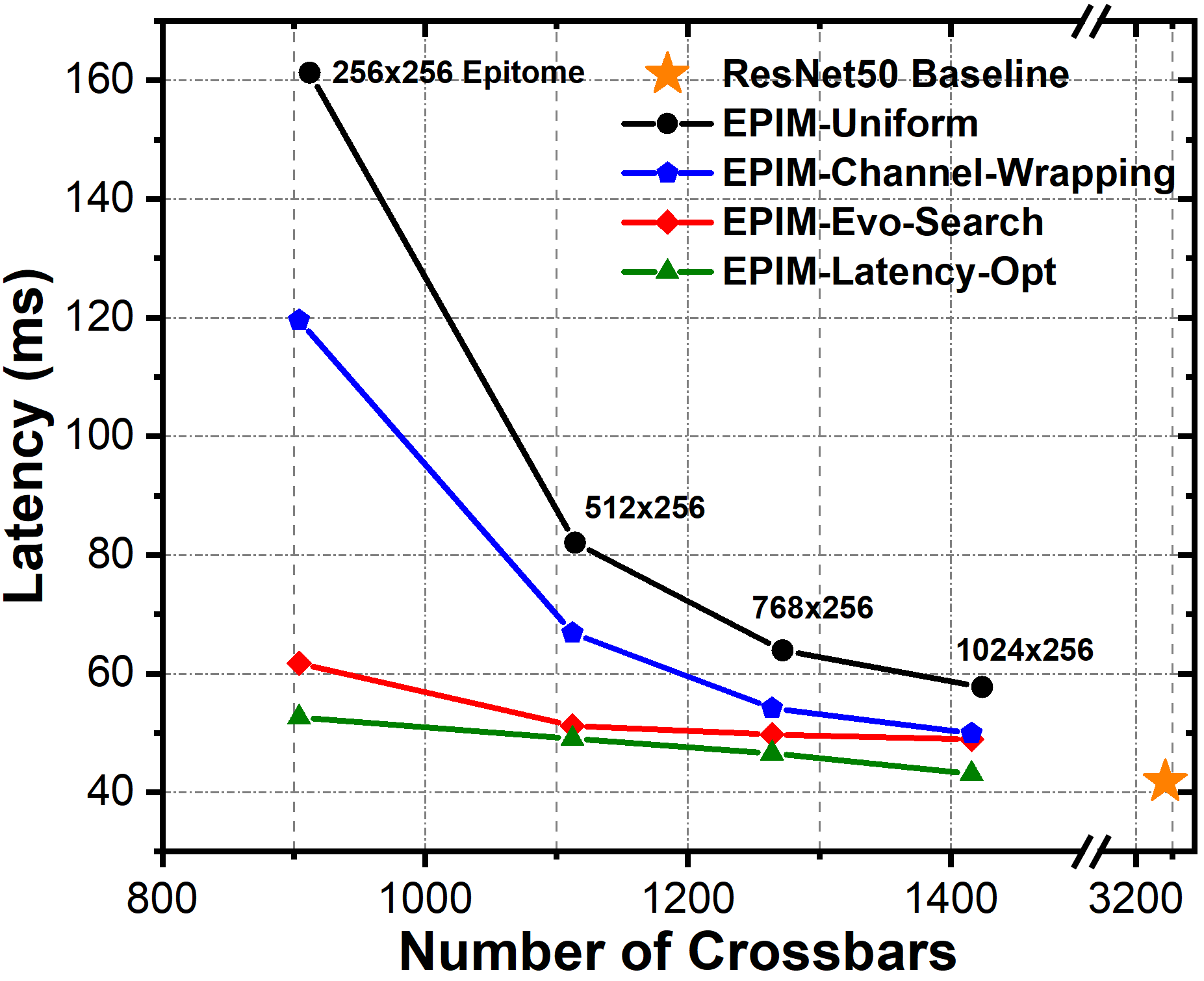}
    \caption{Latency}
    \label{fig:latency_opt}
\end{subfigure}
\hfill
\begin{subfigure}{0.32\textwidth}
    \centering
    \includegraphics[width=0.9\textwidth]{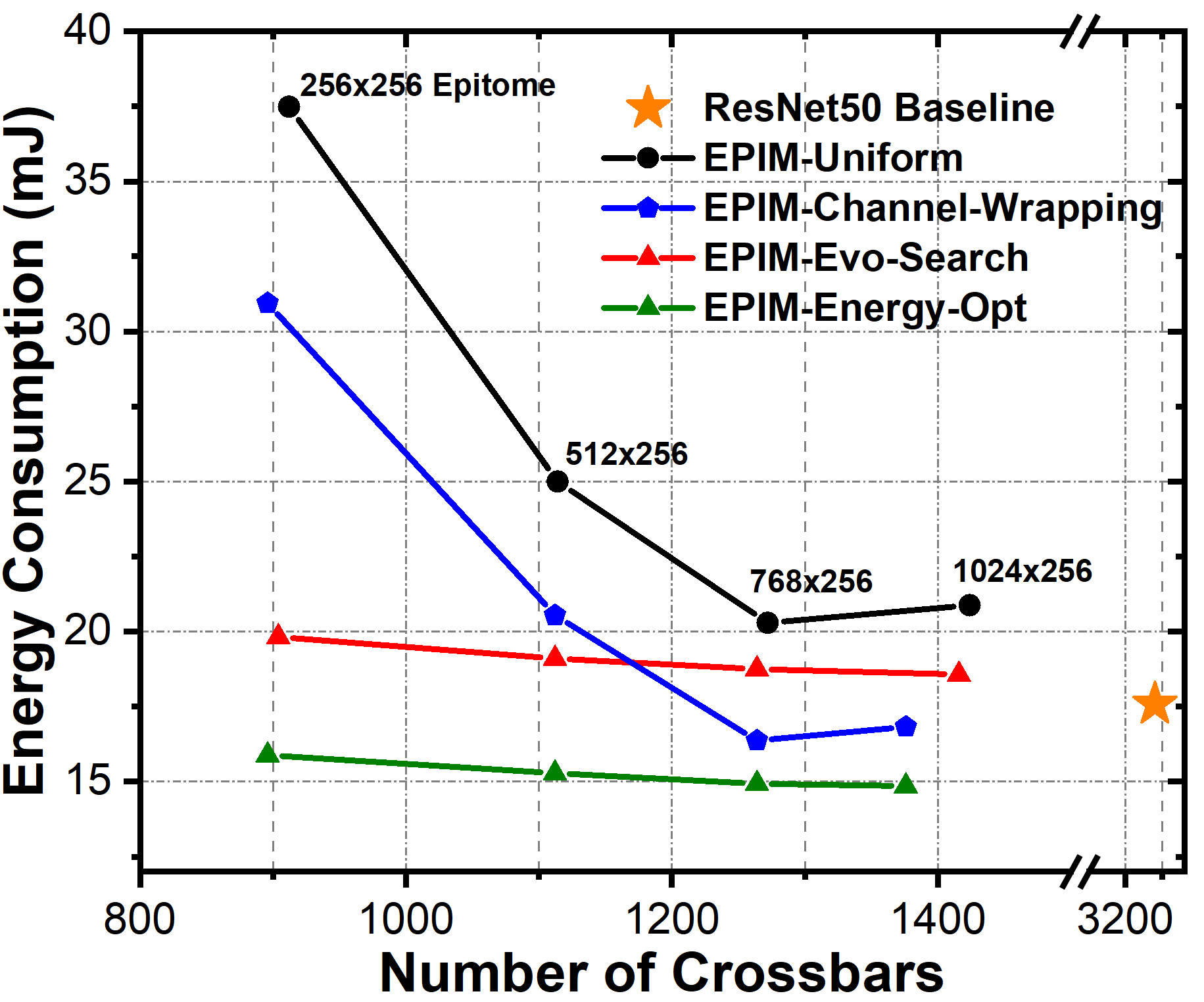}
    \caption{Energy}
    \label{fig:energy_opt}
\end{subfigure}
\hfill
\begin{subfigure}{0.32\textwidth}
    \centering
    \includegraphics[width=0.9\textwidth]{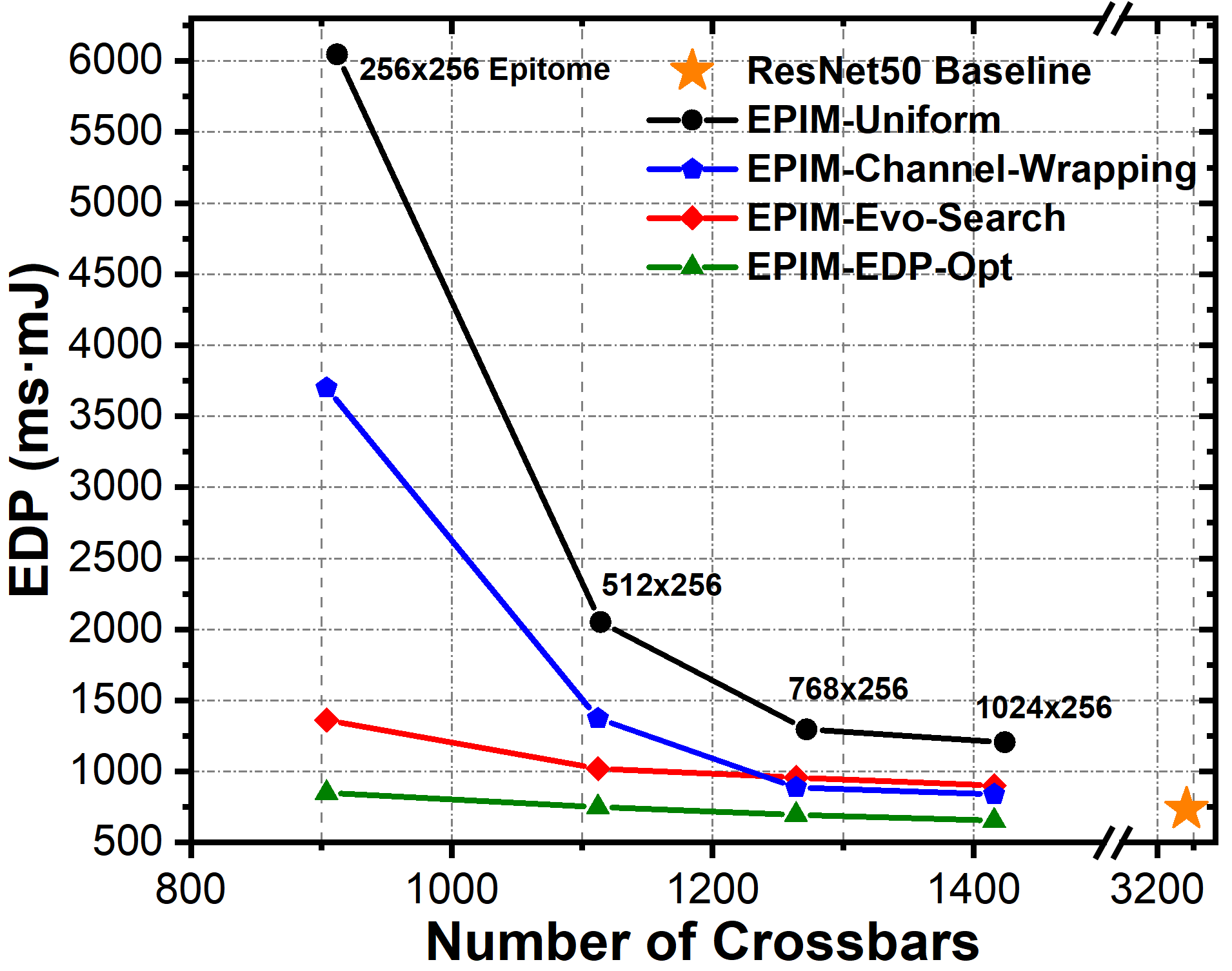}
    \caption{EDP(\underline{E}nergy-\underline{D}elay \underline{P}roduct)}
    \label{fig:edp_opt}
\end{subfigure}
\caption{Comparison of the uniform-epitome solution with two optimization methods: EPIM-Channel Wrapping and EPIM-Evo-Search, each used individually. EPIM-Opt refers to a combination of these two methods.}
\label{fig:comparisons}
\end{figure*}

We evaluated the computational overhead via simulation. Figure \ref{fig:comparisons} indicates that epitomes can heighten both latency and energy consumption. In certain configurations, such as the \textquotedblleft 256x256 Epitome\textquotedblright, the increase is significant, with latency and energy rising to 3.86$\times$ and 2.13$\times$ that of the ResNet-50 Baseline, respectively. 

Upon deeper analysis, we determined that the causes of increased latency and energy consumption differ. 
For latency, due to the repeated activation of memristor crossbars by the epitome, the latency of all hardware components increases proportionally. For example, if we use a epitome with 256 output channels to represent a convolution with 512 output channels, we would need to activate the memristor crossbar at least 2 times. This results in the latency of the epitome increasing proportionally with the number of times the memristor crossbars are activated. As shown in Figure 4(a), the overall latency increase is roughly proportional to the compression rate.

The increase in energy consumption, on the other hand, is a bit more complex. In essence, this increase arises due to the epitome's extensive use of certain high-cost hardware components. Still using the previous example, for this operator, since we need to store a feature map in the buffer each time we activate a small kernel, the output buffer has to be written four times more than for a convolution. As depicted by Figure \ref{fig:energy_opt}, the energy consumption also increases sharply with the compression rate. 

To counteract this increase, we propose layer-wise epitome design and output channel wrapping.

\begin{algorithm}[h]

\small
  \caption{Pseudo code of Evolution Search-based Epitome Design}
  \begin{algorithmic}[1]
    \Require
    Population: $\{P\}_{i}$; Population Filtered by Model Size $\{O\}_{i}$; Evaluation Tools: ET; Evaluation results of candidates: $\{R\}_{i}$
    \State Initialize: $\{P\}_{0}.$init(), i = 0
    \For{$i$ $<$ $Max\ Iteration$}
    \State // Filter population
    \For{each $individual$ in $\{P\}_{i}$}
    \If { Model Size($individual$) $<$ Target Size}
    \State $\{O\}_{i}$.append($individual$)
    \State $\{R\}_{i}$.append(ET($individual$))
    \EndIf
    \EndFor
    \State //Select good candidates:

    \State $Parents$ = sort($\{Reward\}_{i}$, $\{O\}_{i}$)
    \label{line:select}
    \State// Mutate parents
    \For{each $parent$ in $Parents$}
    \State $child$ = mutate($parent, \{P_x\}$)
    \State $\{P\}_{i+1}.$append$(parent)$
    \State $\{P\}_{i+1}.$append$(child)$
    \EndFor
    \EndFor
    \State \Return $\{O\}_{Max\ Iteration}$ 
    \label{line:mutate_end}
    
  \end{algorithmic}
  
  \label{code:recentEnd}
\end{algorithm}

\subsection{Layer-Wise Epitome Design}
\label{subsec:opt_alg}

Within epitome based neural networks, the sensitivity to these increases differs significantly between layers.Figure \ref{fig:opt_motivation} portrays the parameter size, latency, and energy for three layers in ResNet-50. The darker color represents the results using epitomes, while the lighter color stands for the results of convolutions. For instance, layer 67 (depicted in grey) shows that the epitome reduces parameter size by 983.6k, while only introducing a 53.8 ms increase in latency and a 14.5 mJ increase in energy consumption. In contrast, for layer 9 (depicted in red), the epitome reduces the parameter size by merely 20.5k but introduces a 40.8 ms increase in latency and a 10.6 mJ increase in energy consumption.

Considering the varied sensitivities across layers, we propose a tailored epitome design specific to each layer. By creating smaller epitomes for layers that are less impacted by increases, and larger epitomes for those more sensitive, we can substantially lower both latency and energy use while keeping the same compression rate. This process can be written as an optimization problem under constraint as follows.

Let $E$ be a set where its element $e_i$ represents the epitome choice for the $i^{th}$ layer in a total $l$ layer neural network, and $e_i$ is chosen from a set of candidate epitomes $C=\{c_1, c_2, \ldots, c_N\}$. We aim to find the optimal $E^*$ that minimizes the total latency and energy consumption, while using less memristor crossbars as desired level.

However, deciding the optimal $E^*$ poses a significant challenge. Given multiple epitome candidates for each layer, the entire design space boasts $N^l$ potential combinations. This search space can be overwhelmingly vast. In our study, this search space encompasses  20,676,608 epitome combinations. To overcome this challenge, we propose to use evolution search to explore the search space.\par


Algorithm~\ref{code:recentEnd} shows the details of over evolution search algorithm for deciding the optimal epitomes. We use the inverse of latency or energy as reward that the evolution search tries to optimize. To guarantee that the searched epitomes maintains a desired compression rate, we introduce a value $m$ to consider the crossbars used by epitome. We formulate the reward in evolution search as Equation \ref{equ:reward}.
\begin{gather}
\small
\label{equ:reward}
 Reward =  \frac{1}{\text{Latency}(E)} \times m \quad \text{or} \quad \frac{1}{\text{Energy}(E)} \times m, \\
\text{where} \quad  m = \quad 
\begin{cases}
0, &   \text{if \#Crossbar}(E) > \text{Budget}, \\
1, &  \text{if \#Crossbar}(E) \leq \text{Budget}.
\end{cases}
\end{gather}

As is shown in Equation \ref{equ:reward}, if the epitomes $E$ uses more crossbars than desired, the value $m$ will be set as zero.Making the reward lower than any epitome that meets the desire. \par

In each iteration, the evolution search algorithm selects the epitome combination $E$s with the highest rewards as the parent combinations $P$s of the next iteration (\textbf{line} \ref{line:select}). Then, we randomly choose some layers out of the parent combination and randomly set them to other epitomes. The above processes are repeated for many iterations. We use the best epitome combination in the last iteration as $E^*$.

\subsection{Output Channel Wrapping}
\label{subsec:opt_sys}

We can further improve epitome performance by utilizing computational redundancy in the output channel. In our approach, the small patches sampled from epitome are replicated multiple times across output channels. If a patch has $c$ output channels and is used to represent a convolution with $c \times r$ channels, it's duplicated $r$ times. Representing the weight of this virtual convolution as $W$, we can describe the translation invariance property as:
\begin{equation}
\small
W[x,:,:,:] = W[x+c,:,:,:], \quad x = 0,\ldots, c \times (r - 1)
\end{equation}
Such replication prompts a translation invariance in the output feature map as well, which we represent as $OFM$. This characteristic can be formally depicted as:
\begin{equation}
\small
OFM[x,:,:,:] = OFM[x+c,:,:,:], \quad x = 0,\ldots, c \times (r - 1)
\end{equation}

With this property, we can simply calculate $c$ channels of $OFM$ and reuse the result for the rest. In this way, the output buffer write times is reduced by a factor of $r$.  We refer to this method as \textquotedblleft output channel wrapping\textquotedblright.  To support output channel wrapping, we only need to slightly modify IFAT and OFAT by changing the start and stop index of the feature maps. Together with optimization in algorithms, the performance of EPIM will be further improved. An evaluation of the two optimizations is presented in Figure \ref{fig:comparisons}.


\begin{table*}[h]
\centering
\caption{Experimental results of \OURS~on ImageNet. CR stands for compression rate, XB stands for crossbar array, and mp stands for mixed-precision. The dimensions of an epitome are represented by a product notation. For example, \textquotedblleft$1024\times256$\textquotedblright\ represents that $c_{in}\times p \times q = 1024$ and $c_{out} = 256$}
\footnotesize
\renewcommand{\arraystretch}{0.6}
\resizebox{\linewidth}{!}
{
\begin{tabular}{l|c|c|c|c|c|c|c|c}
\toprule
Model &  Bitwidth & Epitome & Accuracy(\%) $\uparrow$ & \# XBs $\downarrow$ & CR of XBs $\uparrow$ & Latency($ms$) $\downarrow$ & Energy($mJ$) $\downarrow$ & \makecell{Memristor \\ Utilization(\%)} $\uparrow$ \\
\midrule
ResNet50         & FP32  &  -                & 76.37   &  13120 & 1.00  & 139.8 & 214.0 & 94.9 \\
\OURS-ResNet50   & FP32  &  1024$\times$256  & 74.00   &  5696  & 2.30  & 167.7 & 194.8 & 96.7 \\
PIM-Prune-ResNet50\cite{chu2020pim}   & FP32  &  -  & 73.18   &  -  & 2.18  & - & - & -\\
\midrule
\OURS-ResNet50   & W9A9    &  1024$\times$256  & 73.98   &  1424  & 9.21  & 50.9  & 17.0 & 96.7 \\
\OURS-ResNet50-Latency-Opt & W9A9 & layer-wise   &   73.60 &  1080  & 12.15  & 49.2  &  16.4 & 93.4 \\
\OURS-ResNet50-Energy-Opt  & W9A9 &  layer-wise   &  73.15 & 1048   & 12.52  &   50.6 & 15.6 & 93.2 \\
\OURS-ResNet50   & W7A9    &  1024$\times$256  & 73.81   &  1076  & 12.19 & 45.2 & 20.5  & 93.2  \\
\OURS-ResNet50   & W5A9    &  1024$\times$256  & 73.59   &  720   & 18.12 & 39.9  & 13.7 & 93.2  \\
\OURS-ResNet50   & W3mpA9   &  1024$\times$256  &  72.98  &  618   & 21.23 & 37.0  &  10.2 & 93.2 \\
\OURS-ResNet50   & W3A9   &  1024$\times$256  &  71.59  &  428   & 30.65 & 36.7  &  9.3  & 93.2\\
\midrule
ResNet101        & FP32  &  -                &  78.77  &  22912 & 1.00  & 189.7 & 385.7 & 94.7\\
\OURS-ResNet101  & FP32  & 1024$\times$256   &  76.56  &  10592 & 2.16  & 263.7 & 364.8 & 98.2 \\
PIM-Prune-ResNet101\cite{chu2020pim}  & FP32  &  -  & 75.82   &  -  & 1.90  & - & - & -\\
\midrule
\OURS-ResNet101  & W9A9    & 1024$\times$256   &  76.52  &  2648  & 8.65  & 75.8  & 32.2 &  98.2 \\
\OURS-ResNet101  & W7A9    & 1024$\times$256   &  76.48  &  1994  & 11.49 & 73.7  & 39.5 & 98.2\\
\OURS-ResNet101  & W5A9    & 1024$\times$256   &  75.68  &  1584  & 14.46 & 72.1  & 29.2 & 98.2  \\
\OURS-ResNet101  & W3mpA9  & 1024$\times$256  &  75.80  &  1052   & 21.78 & 65.5  &  18.6 & 98.2   \\
\OURS-ResNet101  & W3A9   & 1024$\times$256   &  74.98  &  734   & 31.22 & 63.4  &  17.0  & 98.2  \\
\bottomrule
\end{tabular}}

\label{classification}
\end{table*}

\section{Experiments}

\subsection{Experimental Setup}
The development of the simulator is based on MNSIM~\cite{zhu2020mnsim}, a highly accurate, behavior-level simulator.
%
Following the practice in MNSIM, we maintain a look-up table for the storage of the latency and power parameters associated with basic hardware behaviors. The lookup table is modified to include the parameters of epitome.
%
%
%
We use the well-explored 2-bit memristor cells in simulation. For quantization, we integrate the popular HAWQ~\cite{dong2019hawq} algorithm into our simulator. 
All the experiments are conducted on the large-scale benchmark ImageNet~\cite{deng2009imagenet}. 
For comparison purposes, we reproduced PIM-Prune, a pruning framework designed for neural networks on PIM accelerators\cite{chu2020pim}.

\subsection{Results and Analyses}

The main experiment results of EPIM are summarized in Table \ref{classification} and Figure \ref{fig:comparisons}.\par 
Table \ref{classification} provides a comparison between conventional NN algorithms and our EPIM solutions. We choose ResNet-50 and ResNet-101 to demonstrate the experimental results. For ResNet50, our method achieves up to a $30.65\times$ compression rate of crossbars with a maximum accuracy loss of $4.78\%$. For ResNet-101, our method achieves up to a $31.22\times$ compression rate of crossbars with a maximum accuracy loss of $3.79\%$. More importantly, combined with ultra-low bit quantization, EPIM reduces the inference energy by 23.01 $\times$ for ResNet-50 and 22.64$\times$ for ResNet-101. Our experiments show that EPIM can serve as a competitive solution to achieve efficient PIM deployment of large CNNs.\par


In Figure~\ref{fig:comparisons}, we contrast manually designed uniform epitomes with epitomes produced by design optimization methods. The EPIM-Opt method, combining Channel Wrapping and Evo-Search, outperforms others. Compared to uniform epitomes, it offers similar compression with up to 3.07× speedup, 2.36× energy savings, and 7.13× reduced EDP. A comparison of ResNet50 under W9A9 quantization in Table \ref{classification} reveals that EPIM-ResNet50-Latency-Opt exhibits the lowest latency, while EPIM-ResNet50-Energy-Opt presents the lowest energy consumption, both without significant accuracy degradation.

\vspace{-5mm}
\section{Ablation Study}

\subsection{Detailed Experiments for Quantization}
The detailed results of quantization are provided in Table \ref{tab:quantization_results}. We retrained six quantized models under two ultra-low-bit quantization scenarios. The results show that both of the proposed updates contribute to the increase in accuracy. For instance, for ResNet-50 under 3-bit quantization, the accuracy can be improved from 69.95\% to 71.35\% by adjusting scaling factor with crossbars. Moreover, by using the weighted sum scaling factor, the accuracy can be further improved to 71.59\%.

\begin{table}[h]
\centering
\caption{Detailed Quantization Experiments Results.}
\footnotesize  
\renewcommand{\arraystretch}{0.6}  
\begin{tabular}{l|c|c|c}
\toprule
& \multicolumn{3}{c}{Accuracy (\%)} \\
\makecell{Model} & \makecell{Naïve\\ Quant} & \makecell{+ Adjust with \\ Crossbars} & \makecell{+ Adjusted with \\Overlap} \\
\midrule
ResNet-50 (3-bit) & 69.95 & 71.35 & 71.59 \\
\midrule
ResNet-50 (3-5 bit) & 72.18 & 72.83 & 72.98 \\
\midrule
ResNet-101 (3-bit) & 73.98 & 74.96 & 74.98 \\
\midrule
ResNet-101 (3-5 bit) & 75.46 & 75.71 & 75.80 \\
\bottomrule
\end{tabular}

\label{tab:quantization_results}
\end{table}

\vspace{-3mm}
\subsection{Detailed Comparison of Epitome and Pruning}

In order to assess the epitome's compatibility with pruning, we experiment with merging both concepts using basic element-wise pruning methods. Due to challenges in determining the crossbar compression rate with pruning, we compare parameter compression rates. Results are in Table \ref{tab:pruning_results}.

\begin{table}[h]
\centering
\caption{Accuracy and compression rate of epitome, epitome + element pruning with 50\% pruning ratio, PIM-Prune\cite{chu2020pim}.}
\footnotesize  
\renewcommand{\arraystretch}{0.6}  
\begin{tabular}{l|c|c|c|c}

\toprule
& \multicolumn{2}{c|}{ResNet-50} & \multicolumn{2}{c}{ResNet-101} \\
Method & \makecell{Accuracy\\ (\%)} & \makecell{Compress.\\ Rate} & \makecell{Accuracy\\ (\%)} & \makecell{Compress.\\ Rate} \\
\midrule
Epitome & 74.00 & 2.25 & 76.56 & 2.08 \\
\midrule
Epitome + Pruning & 73.18 & 3.49 & 75.76 & 3.64 \\
\midrule
PIM-Prune 50\% & 72.77 & 1.80 & 75.82 & 1.90 \\
\midrule
PIM-Prune 75\% & 72.19  & 3.38 & 74.80 & 3.24 \\
\bottomrule

\end{tabular}
\vspace{-3mm}
\label{tab:pruning_results}
\end{table}
\vspace{-0mm}

\section{Conclusion}
In this paper, we introduce the epitome for model compression in PIM accelerators. We redesign the PIM architecture and utilize epitome-aware quantization, layer-wise epitome design and output channel wrapping to enhance performance. Our \OURS\ method achieves a 71.59\% top-1 accuracy on ImageNet using 3-bit \OURS-ResNet50, reducing crossbar areas by 30.65$\times$.


\bibliographystyle{ACM-Reference-Format}
\bibliography{main.bib}

\end{document}